\documentclass[a4paper]{jpconf}

\usepackage{graphicx}
\usepackage{latexsym,amssymb,amsmath,amsfonts}

\def\id{\makebox[0.6ex][l]{$1$}{\rm l}} 

\begin{document}

\title{Anomalous Lorentz and CPT violation\footnote{Invited talk at the
\textit{International Workshop on CPT and Lorentz Symmetry in Field Theory}, University of the Algarve, Faro, Portugal, July 6--7, 2017.
\hfill (J. Phys. Conf. Ser. \textbf{952} (2018) 012003; arXiv:1709.01004)}
}

\author{F R Klinkhamer}

\address{Institute for Theoretical Physics,
Karlsruhe Institute of Technology (KIT), 76128 Karlsruhe, Germany}

\ead{frans.klinkhamer@kit.edu}

\begin{abstract}
If there exists Lorentz and CPT violation in nature, then it is crucial
to discover and understand the underlying mechanism.
In this contribution, we discuss one such mechanism which relies
on four-dimensional chiral gauge theories defined over
a spacetime manifold with topology $\mathbb{R}^3 \times S^1$
and periodic spin structure for the compact dimension.
It can be shown that the effective gauge-field action contains
a local Chern-Simons-like term which violates Lorentz and CPT invariance.
For arbitrary Abelian $U(1)$ gauge fields with trivial holonomies
in the compact direction, this anomalous Lorentz and CPT violation
has recently been established perturbatively with a Pauli--Villars-type regularization and nonperturbatively with a lattice
regularization based on Ginsparg-Wilson fermions.
\end{abstract}

\thispagestyle{plain}\pagenumbering{arabic}  

\newpage
\section{Introduction}
\label{sec:Introduction}

Experiment has shown the violation of P, C, CP, and T,
but \emph{not} of CPT.
Indeed, there is the well-known CPT ``theorem''
(L\"uders, 1954--57; Pauli, 1955; Bell, 1955;  Jost, 1957),
which states that
any local relativistic quantum field theory is invariant under the combined operation (in whichever order)
of charge conjugation (C), parity reflection (P), and
time reversal (T).
The formulation of the theorem can be sharpened, but
the form stated suffices for the moment.

The main inputs of the theorem are
\begin{itemize}
\item flat Minkowski spacetime
      $(M,\, g_{\mu\nu}) = \left(\,\mathbb{R}^4,\, \eta_{\mu\nu}^\text{Minkowski}\,\right)$;
\item invariance under  proper or\-tho\-chro\-nous Lorentz transformations
      and spacetime translations;
\item standard spin-statistics connection;
\item locality and Hermiticity of the Hamiltonian.
\end{itemize}
We refer to two monographs~\cite{Sakurai1964,StreaterWightman1964}
for further details and references on the CPT theorem.

The following question arises:
\emph{can}  CPT invariance be violated at all
in a physical theory and, if so, \emph{is} it in the real world?
It was widely believed that only quantum-gravity
or superstring  effects could give CPT violation.
But a different result has been obtained several years ago:
for certain spacetime topologies and  classes of
chiral gauge theories, CPT invariance is broken anomalously,
that is, by quantum effects.

The original paper for this ``CPT anomaly'' is
Ref.~\cite{Klinkhamer2000}.
Follow-up papers have appeared in
Refs.~\cite{KlinkhamerNishimura2001,KlinkhamerMayer2001,KlinkhamerSchimmel2002,%
KlinkhamerRupp2004,GhoshKlinkhamer2017}
and an extensive review is given by Ref.~\cite{Klinkhamer2005}.
The crucial ingredients of the CPT anomaly are
\begin{itemize}
\item chiral fermions and gauge interactions;
\item non-simply-connected spacetime topology,
\item periodic boundary conditions.
\end{itemize}
Expanding on the last two bullets,
the spacetime manifold must have a separable compact
spatial dimension (coordinate $x^3$) with periodic spin structure
of the fermions. The main focus, up till now, has been on the
topology $\mathbb{R}^3 \times S^1$, but also other topologies
have been considered, for example topologies related to
punctures or wormholes~\cite{KlinkhamerRupp2004}.

There are, at least, three possible applications of the CPT anomaly.
First, there is the resulting optical activity of the vacuum,
which may lead to observable effects for the Cosmic Microwave Background~\cite{Klinkhamer2005,CarrollFieldJackiw1990,WMAP2009}.
Second, there is the resulting
fundamental arrow-of-time, which may play a role
in explaining the ``start'' of the Big Bang~\cite{Klinkhamer2002}.
Third, the CPT anomaly may also be used as a
diagnostic tool for a hypothetical spacetime
foam~\cite{KlinkhamerRupp2004,KlinkhamerRupp2005}.

In this short contribution, we focus on the basic mechanism of
the CPT anomaly and skip possible applications
(referring to the review~\cite{Klinkhamer2005}
for further discussion of the phenomenology).

\section{Heuristics}
\label{sec:Heuristics}

The CPT anomaly of a chiral gauge theory defined over
the four-dimensional manifold \mbox{$M=\mathbb{R}^3 \times S^1$,}
with trivial vierbeins $e^a_\mu (x)= \delta^a_\mu\,$
and appropriate background gauge fields,
arises in four steps:
\begin{itemize}
\item a compact spatial dimension with coordinate $x^3 \in [0,L]$ and
      a periodic spin structure make that a chiral fermion can have a
      momentum component $p_3 =0\,$;
\item a single chiral 2-component Weyl fermion
      in four dimensions (4D) with $p_3 =0$
      corresponds to a single massless 2-component Dirac fermion in
      three dimensions (3D)\,;
\item a single massless Dirac fermion in 3D is known to have
      a ``parity anomaly,'' provided gauge invariance is maintained
      exactly~\cite{Redlich1984,AlvarezGaume-etal1985,CosteLuescher1989}\,;
\item this ``parity'' violation in 3D corresponds to T violation in 4D,
      which, in turn, implies CPT violation in 4D\,.
\end{itemize}

The actual CPT anomaly will be established by calculating the
effective gauge-field action $\Gamma[A]$, where the effects
of the virtual fermions have been integrated out,
and by showing that this effective action $\Gamma[A]$ changes
under a CPT transformation of the background gauge field $A$.

\section{Perturbative calculation}
\label{sec:Perturbative-calculation}

\subsection{Main result}
\label{subsec:Main-result-pert}

Consider, for definiteness, a chiral gauge theory with
the following gauge group $G$,
left-handed-fermion representation $R_L$, and spacetime manifold $M$:
\begin{subequations}\label{eq:G-RL-M-NonAbelian}
\begin{eqnarray}
\label{eq:G-NonAbelian}
G &=& SO(10)\,,
\\[1.5mm]
\label{eq:RL-NonAbelian}
R_L&=&N_\text{fam}\times (\mathbf{16})\,,
\quad
N_\text{fam}=1\,,
\\[1.5mm]
\label{eq:M-NonAbelian}
M\!\! &=&\!\!\mathbb{R}^3 \times S^1_\text{\,PSS}\,,
\quad
e^a_\mu (x)= \delta^a_\mu\,,
\quad
g_{\mu\nu}(x)= e^a_\mu (x) \, e^b_\nu (x)\,\eta_{ab}=\eta_{\mu\nu} \,,
\end{eqnarray}
\end{subequations}
where the subscript
``PSS'' in \eqref{eq:M-NonAbelian} stands for periodic spin structure
and the metric $g_{\mu\nu}(x)$ has a Lorentzian signature $(-1,1,1,1)$.
The spacetime manifold $M$ from \eqref{eq:M-NonAbelian}
is described by the following coordinates $x^\mu$:
\begin{eqnarray}\label{eq:x012range-x3range}
x^0 , x^1, x^2 &\in& \mathbb{R}  \,,
\quad
x^3 \in [0 , L]  \,.
\end{eqnarray}
The spinorial $SO(10)$ representation $\mathbf{16}$
from \eqref{eq:RL-NonAbelian}
is a complex representation (see also the remarks in the
first paragraph of Sec.~\ref{subsec:Main-result-nonpert}).
For simplicity,
we have set $N_\text{fam}$ in \eqref{eq:RL-NonAbelian}
equal to $1$, but similar results hold for the value $N_\text{fam}=3$
relevant to elementary particle physics.

The fermionic field in the representation \eqref{eq:RL-NonAbelian}
is denoted by $\psi_L(x)$
and the Lie-algebra-valued gauge field $A_{\mu}(x)$
is defined as follows:
\begin{subequations}\label{eq:YM-field-Ta-normalization}
\begin{eqnarray}\label{eq:YM-field}
A_{\mu}(x) &\equiv& g\,A_{\mu}^{a}(x)\,T^{a}\,,
\\[1.5mm]
\label{eq:Ta-normalization}
\text{tr}\, \Big( T^{a}\,T^{b}\Big) &=&
- \frac{1}{2} \;\delta^{ab}\,,
\end{eqnarray}
\end{subequations}
with
argument $x$ standing for  $(x^0,\,x^1,\, x^2,\, x^3)$,
Yang--Mills coupling constant $g$,
and real gauge fields $A_{\mu}^{a}(x)$.
The index ``a'' in \eqref{eq:YM-field} labels the
anti-Hermitian generators $T^{a}$ of the $\text{so}(10)$ Lie algebra
and is summed over ($a=1,\, \ldots \,, 45$),
while \eqref{eq:Ta-normalization}
specifies the normalization of the generators.

Both gauge and fermionic fields are assumed to have
periodic boundary conditions in $x^3$,
\begin{subequations}\label{eq:periodic-bcs}
\begin{eqnarray}
\label{eq:periodic-bcs-A}
A_\mu(\widetilde{x},\, x^3+L)
&=&
A_\mu(\widetilde{x},\, x^3)   \,,
\\[1.5mm]
\label{eq:periodic-bcs-psiL}
\psi_L(\widetilde{x},\, x^3+L)
&=&
\psi_L(\widetilde{x},\, x^3)  \,,
\end{eqnarray}
\end{subequations}
with $\widetilde{x}\equiv (x^0,\,x^1,\, x^2)$.
The coupling between gauge and fermionic fields is specified
by the following Dirac equation
(equivalent to the standard Weyl equation \cite{ItzyksonZuber1980}
by use of the chiral representation of the Dirac matrices $\gamma^\mu$):
\begin{equation}
\label{eq:Dirac-equation-non-Abelian}
\gamma^\mu\,\Big(\partial_\mu + A_\mu(x)\Big)\,\psi_L(x)=0\,,
\end{equation}
where the coupling constant $g$ has been absorbed in
the gauge field according to \eqref{eq:YM-field}.

For this setup, the complete effective gauge-field action
$\Gamma[A]$ for  $A$ $\in$ $\text{so}(10)$
is, of course, not known exactly
(there are certain exact results
in 2D~\cite{KlinkhamerNishimura2001,KlinkhamerMayer2001}).
But the crucial term in $\Gamma[A]$ has been identified
perturbatively~\cite{Klinkhamer2000,GhoshKlinkhamer2017}
for a gauge field with trivial holonomies (e.g., $A_{3}=0$):
\begin{eqnarray}
\label{eq:CS-like-term}
\Gamma_\text{\,anom}^{\,\mathbb{R}^3 \times S^1}[A\,]
&\!\!=\!\!&
 \frac{1}{16\,\pi^2} \;
   \int_{\mathbb{R}^3} dx^0 dx^1 dx^2
   \int_{0}^{L} dx^3\;
\frac{\pi}{L}\;\epsilon^{\kappa\lambda\mu 3}
 \;\text{tr}\, \left(
 A_{\kappa}(x)\, A_{\lambda\mu}(x)
 -\frac{2}{3}\, A_{\kappa}(x)\, A_{\lambda}(x)\, A_{\mu}(x)
 \right) \nonumber\\[1mm]
 &&
 + \ldots\,,
\end{eqnarray}
in terms of the Levi-Civita symbol $\epsilon^{\kappa\lambda\mu\nu}$,
the Lie-algebra-valued gauge field $A_{\kappa}(x)$ from
\eqref{eq:YM-field},
and the corresponding Yang--Mills field strength $A_{\lambda\mu}(x)$,
\begin{eqnarray}
\label{eq:YM-field-strength}
A_{\lambda\mu}(x) &\equiv&
\partial_{\lambda} A_{\mu}(x)
-\partial_{\mu} A_{\lambda}(x)
+A_{\lambda}(x)\,A_{\mu}(x)
-A_{\mu}(x)\,A_{\lambda}(x)\,.
\end{eqnarray}
The ellipsis in \eqref{eq:CS-like-term} contains further (nonlocal) terms;
see the third remark in Sec.~\ref{subsec:Technical-remarks}.

The result \eqref{eq:CS-like-term}
has been obtained perturbatively at the one-loop level
for arbitrary gauge fields with trivial holonomies,
\begin{equation}
\label{eq:holonomy-condition}
\forall \widetilde{x} \in \mathbb{R}^3\,:\;
H_3 (\widetilde{x})=\id \,,
\end{equation}
where the holonomy along the straight path in the 3 direction
starting at $\widetilde{x}\equiv (x^0,\,x^1,\, x^2)$
and $x^3=0$ is defined by
\begin{equation}
\label{eq:holonomy-nonAbelian}
H_3 (\widetilde{x})
\equiv
\mathcal{P}\, \exp \left[ \int_0^{L} dy\, A_3(\widetilde{x},\,y) \right]
\in G \,,
\end{equation}
with the path-ordering operator $\mathcal{P}$.
Taking into account the periodic
boundary conditions \eqref{eq:periodic-bcs-A},
the trace of the holonomy $H_3 (\widetilde{x})$
corresponds to a gauge-invariant quantity,
also known as the Wilson loop variable.

The spacetime integral in \eqref{eq:CS-like-term}
involves the Chern-Simons density~\cite{ChernSimons1974},
\begin{equation}
\label{eq:omega-CS}
\omega_\text{CS}[A_0,\,A_1,\,A_2]\equiv
\frac{1}{16\,\pi^2} \,\epsilon^{klm} \;
\text{tr}\,\left(A_{k}\,A_{lm} -\frac{2}{3}\,A_{k}\,A_{l}\,A_{m}\right)\,,
\end{equation}
where the repeated indices $k$, $l$, and $m$
are summed over $\{0,\,1,\,2 \}$.
A genuine topological Chern--Simons term
$\Omega_\text{CS}=\int \omega_\text{CS}$
is obtained only for a 3-dimensional spacetime
manifold~\cite{ChernSimons1974}.
Our anomalous action term \eqref{eq:CS-like-term} holds, however,
in four spacetime dimensions:
the integration is over four spacetime coordinates
and the gauge fields $A_\mu$ also have a dependence on $x^3$.
Hence, the qualification ``Chern--Simons-like'' for the
anomalous local action term in \eqref{eq:CS-like-term}.
This local action term is nontopological
in the sense that there is a nontrivial dependence
on the spacetime metric or vierbein;
see, e.g., Ref.~\cite{KantKlinkhamer2005}
and also the further remarks in the last paragraph of Sec.~5.

The local term \eqref{eq:CS-like-term}
is \emph{Lorentz-noninvariant},
because of the explicit spacetime index ``3'' entering the
Levi--Civita symbol, and is also \emph{CPT-odd},
because of the odd number (namely, three) of spacetime indices
for the gauge-field terms in the large brackets. Recall that
the standard Yang--Mills action density
term $\text{tr}\,\big(A_{\mu\nu}(x)\,A^{\mu\nu}(x)\big)$
is Lorentz-invariant and CPT-even~\cite{ItzyksonZuber1980}.

\subsection{Technical remarks}
\label{subsec:Technical-remarks}

In this subsection, we present six technical remarks, which
can, however, be skipped in a first reading.
First, the non-Abelian result \eqref{eq:CS-like-term}
also holds for an Abelian $U(1)$ group,
as long as attention is given to the proper normalization.
Indeed, the single Lie algebra generator $T^1$ can
be written as $T^1= \big(i/\sqrt{2n}\,\big)\,\id_n$\,, in terms of
an $n \times n$ identity matrix $\id_n$. Now take $n=1$, so that
the trace operation in \eqref{eq:CS-like-term} becomes trivial.
The covariant derivative \eqref{eq:Dirac-equation-non-Abelian}
on a single left-handed fermion $\chi_L$ then reads
$\big(\partial_\mu + i\,(g/\sqrt{2})\, A^1_\mu\big)\,\chi_L$
and we can define
the Abelian $U(1)$ gauge field $B_\mu(x)\equiv A^1_\mu(x)$
and the $U(1)$ charge $e\equiv g/\sqrt{2}$.
This rather explicit discussion aims to clarify some
confusing statements in Ref.~\cite{Klinkhamer2000},
where the parameter ``$a$'' in Eqs.~(3.13) and (4.1)
is simply to be omitted.

Second,
the following expression is obtained by
writing $\epsilon^{\kappa\lambda\mu3}$ in the integrand
of \eqref{eq:CS-like-term}
as $\epsilon^{\kappa\lambda\mu\nu}\,\partial_\nu x^3$
and performing a partial integration (for gauge fields
vanishing at infinity):
\begin{equation}
\label{eq:Gamma-anom}
\Gamma_\text{\,anom}^{\,\mathbb{R}^3 \times S^1}[A\,] =
 \frac{1}{32\,\pi} \;
   \int_{\mathbb{R}^3} dx^0 dx^1 dx^2
   \int_{0}^{L} dx^3\;  \frac{x^3}{L}\;
   \epsilon^{\kappa\lambda\mu\nu}
 \;\text{tr}\, \Big( A_{\kappa\lambda}(x)\, A_{\mu\nu}(x) \Big)
 + \ldots \,,
\end{equation}
where the integrand depends solely on
the Yang--Mills field strength  \eqref{eq:YM-field-strength}.
The spacetime-dependent ``coupling constant'' $x^3/L$ in
\eqref{eq:Gamma-anom} makes clear that Lorentz invariance is broken.

Third, let us briefly discuss the issue of non-Abelian gauge invariance
(further details can be found in Sec.~4 of Ref.~\cite{Klinkhamer2000}).
The integrand of the
non-Abelian Chern--Simons-like term in \eqref{eq:CS-like-term}
is not completely gauge invariant
(invariant only under infinitesimal 4D gauge transformations)
but is manifestly periodic in $x^3$
because of the boundary conditions \eqref{eq:periodic-bcs-A}.
The integrand of the
non-Abelian term \eqref{eq:Gamma-anom} is completely gauge invariant
(invariant also under ``large'' 3D gauge transformations
and infinitesimal 4D deformations thereof)
but is not manifestly periodic in $x^3$
(in addition, the single coordinate $x^3$ is not a ``good'' coordinate
and two patches are needed to cover the circle $S^1$).
Incidentally, the nonlocal terms of the ellipsis in
\eqref{eq:CS-like-term} are precisely there to restore
the full non-Abelian gauge invariance
(cf. Refs.~\cite{AlvarezGaume-etal1985,CosteLuescher1989}).

Fourth,
the anomalous Chern--Simons-like term
\eqref{eq:CS-like-term} for non-Abelian gauge fields has,
strictly speaking, only been derived~\cite{Klinkhamer2000}
for a special class of background fields,
namely \mbox{$x^3$-independent} gauge fields
with vanishing component in the compact direction,
\begin{subequations}\label{eq:special-class-A}
\begin{eqnarray}
A_{3}&=&0  \,,
\\[1.5mm]
A_{\widetilde{\mu}}&=&A_{\widetilde{\mu}}(\widetilde{x})\,,
\end{eqnarray}
\end{subequations}
with definition $\widetilde{x}\equiv (x^0,\,x^1,\,x^2)$
and $\widetilde{\mu}$ taking values from the set $\{ 0,\,1,\,2\}$.

Fifth,
it is possible to consider a larger class of background fields,
namely $x^3$-dependent Abelian $U(1)$ gauge fields
$B_\mu(x) \in \mathbb{R}$
with vanishing component in the compact direction,%
\begin{subequations}\label{eq:larger-class-B}
\begin{eqnarray}
\label{eq:larger-class-B-B3}
B_{3}&=&0\,,
\\[1.5mm]
\label{eq:larger-class-B-B012}
B_{\widetilde{\mu}}&=&B_{\widetilde{\mu}}(\widetilde{x},\,x^3)\,,
\\[1.5mm]
\label{eq:eq:larger-class-B-periodic-bcs-B}
B_{\widetilde{\mu}}(\widetilde{x},\, x^3)
&=&
B_{\widetilde{\mu}}(\widetilde{x},\, x^3+L)\,,
\end{eqnarray}
\end{subequations}
together with periodic boundary conditions \eqref{eq:periodic-bcs-psiL}
on the relevant fermionic fields $\chi_L(x)$.
For Abelian $U(1)$ gauge fields \eqref{eq:larger-class-B}
with definitions from the first remark of this subsection and with a
Pauli--Villars-type regularization~\cite{FrolovSlavnov1993},
the Abelian version of the local Chern--Simons-like term in \eqref{eq:CS-like-term} has been obtained~\cite{GhoshKlinkhamer2017},
\begin{equation}
\label{eq:CS-like-term-Abelian}
\Gamma_\text{\,anom}^{\,\mathbb{R}^3 \times S^1}[B\,] =
 \frac{1}{16\,\pi^2} \;F\,e^2 \;
   \int_{\mathbb{R}^3} dx^0 dx^1 dx^2
   \int_{0}^{L} dx^3\;
\frac{2\,\pi}{L}\;\epsilon^{\kappa\lambda\mu 3} \;
 B_{\kappa}(x)\,\partial_{\lambda} B_\mu(x) + \ldots \,.
\end{equation}
The numerical factor $F$ in \eqref{eq:CS-like-term-Abelian}
counts the sum of the squares of the fermion
charge in units \mbox{of $e$;} see also Sec.~\ref{subsec:Continuum-limit}.
The derivation of \eqref{eq:CS-like-term-Abelian} requires that
the typical momenta of the background fields \eqref{eq:larger-class-B}
are very much less than the regulator masses.

Sixth,
the anomalous term \eqref{eq:CS-like-term} can be generalized
by including an additional
prefactor $\big(2\,k^{(0)}+1\big)$ for $k^{(0)}\in \mathbb{Z}$,
due to the freedom in defining the regularized theory.
This freedom is particularly clear in the lattice formulation
to be discussed in the next section, where the fermion measure
of the path integral is characterized by an integer $k^{(0)}$.

\section{Nonperturbative calculation}
\label{sec:Nonperturbative-calculation}

\subsection{Main result}
\label{subsec:Main-result-nonpert}

Consider, for definiteness, a four-dimensional Abelian
chiral gauge theory with
\begin{subequations}\label{eq:G-RL-Abelian}
\begin{eqnarray}
\label{eq:G-Abelian}
G &=& U(1), \;
\\[1.5mm]
\label{eq:RL-Abelian}
R_L
&=&
6\times \left(\frac{1}{3}\right) +
3\times \left(-\frac{4}{3}\right) + 3\times \left(\frac{2}{3}\right) +
 2\times \big(-1\big) + 1\times \big(2\big) + 1\times \big(0\big)\; .
\end{eqnarray}
\end{subequations}
The perturbative chiral gauge anomalies of this particular
theory cancel out, because the group \eqref{eq:G-Abelian}
is a subgroup of the ``safe'' group $SO(10)$
and the reducible representation \eqref{eq:RL-Abelian}
of $U(1)$ is contained in the
irreducible representation $\mathbf{16}$  of $SO(10)$.
Observe that $R_L$ from \eqref{eq:RL-Abelian} is a complex
representation: there are, for example, six left-handed fermions
with a normalized charge $+1/3$ but
none with a normalized charge $-1/3$.
This may be compared to the vectorlike gauge theory
of quantum electrodynamics
with the real representation $R_L=(+1)+(-1)$.

Next, define a chiral lattice gauge theory
over a finite hypercubic lattice with
\vspace{0.25\baselineskip}
\begin{itemize}
\item periodic spin structure in one direction;
\item Ginsparg--Wilson fermions~\cite{GinspargWilson1982};
\item Neuberger's lattice Dirac operator~\cite{Neuberger1998a,Neuberger1998b};
\item L{\"u}scher's chiral constraints~\cite{Luescher1998,Luescher1999}.
\end{itemize}
The fermionic fields are, as usual, associated with the
lattice sites and the gauge fields with the directed links between
neighbouring lattice sites. Specifically, the fields are denoted
by $\psi_L(x)$ at the lattice site $x$
and $U_{\mu}(x) \in G$ for the link in the $\mu$ direction
starting from the lattice site $x$.

The Euclidean effective gauge-field
action $\Gamma[U]$ is given by the Euclidean path integral
over the fermionic fields $\psi_L(x)$.
The goal is to establish that this effective  gauge-field
action $\Gamma[U]$ changes under a CPT transformation,
\begin{equation}
\label{eq:CPT-nonivariance-GammaU}
\Gamma[U] \neq \Gamma[U^\text{CPT}]\,,
\end{equation}
where $U$ denotes the set of link variables of the lattice.

The result \eqref{eq:CPT-nonivariance-GammaU} has been established nonperturbatively~\cite{KlinkhamerSchimmel2002,GhoshKlinkhamer2017}
for Abelian $U(1)$ gauge fields with trivial holonomies
(e.g., $U_{3}= 1$).
Moreover, the origin of the CPT anomaly has been identified
as an ambiguity~\cite{KlinkhamerSchimmel2002} in
the choice of basis vectors for the fermion integration
measure; cf. the path-integral derivation of the triangle
(Adler--Bell--Jackiw) anomaly~\cite{Fujikawa1980}.

The result \eqref{eq:CPT-nonivariance-GammaU} has, in particular,
been obtained~\cite{GhoshKlinkhamer2017}
for $U(1)$ link variables that are
\mbox{$x^3$-dependent} and trivial in the compact direction,
\begin{subequations}\label{eq:larger-class-U}
\begin{eqnarray}
U_{3}&=&1 \,,
\\[1.5mm]
U_{\widetilde{\mu}}&=&U_{\widetilde{\mu}}(\widetilde{x},\,x^3)\,,
\\[1.5mm]
U_{\widetilde{\mu}}(\widetilde{x},\, x^3)
&=&
U_{\widetilde{\mu}}(\widetilde{x},\, x^3+L)\,,
\end{eqnarray}
\end{subequations}
with definition $\widetilde{x}\equiv (x^0,\,x^1,\,x^2)$
and $\widetilde{\mu}$ taking values from the set $\{ 0,\,1,\,2\}$.
All arguments in \eqref{eq:larger-class-U}
are understood to coincide with the points of the four-dimensional
hypercubic lattice.

\subsection{Continuum limit}
\label{subsec:Continuum-limit}

In addition to establishing the existence of the
CPT anomaly nonperturbatively, we have also investigated
the continuum limit, with lattice spacing $a\to 0$
and number of links in the 3 direction $N_3\to\infty$,
while keeping $L\equiv N_3\,a$ at a constant value.
For slowly-varying link variables
$U_{\mu}(x) \approx \exp[i \,e\, a\, B_{\mu}(x)]$
in the  continuum limit,
the anomalous term $\Gamma_\text{anom}[U]$ from the
nonperturbative lattice calculation
has been shown~\cite{GhoshKlinkhamer2017} to give rise to the
Abelian Chern--Simons-like term \eqref{eq:CS-like-term-Abelian}
with a numerical factor
$F\equiv \sum_{f}\,(q_f/e)^2=40/3$ for the
representation \eqref{eq:RL-Abelian}.
As mentioned in the last remark of Sec.~\ref{subsec:Technical-remarks},
there is an additional prefactor $\big(2\,k^{(0)}+1\big)$
coming from the definition of the path-integral fermion measure
characterized by an integer $k^{(0)}$
(see App.~B of Ref.~\cite{KlinkhamerSchimmel2002}).

Hence, the nonperturbative lattice result does not modify
the result of the perturbative one-loop calculation
as discussed in Sec.~\ref{sec:Perturbative-calculation}.
This is certainly reminiscent of the
Adler--Bardeen result for the triangle anomaly~\cite{AdlerBardeen1969},
but it remains to be confirmed that there arise no additional
terms in the nonperturbative lattice calculation.

\section{Outlook}
\label{sec:Outlook}

The subtle role of topology for the local properties
of quantum field theory
is well-known, the prime example being the Casimir effect.
For certain chiral gauge theories, we now have established that
the interplay of ultraviolet and infrared
effects may also lead to Lorentz and CPT noninvariance,
even for flat spacetime manifolds, that is, without local gravity.

Let us end with two general remarks.
First, we have found an anomalous origin of
Lorentz and CPT violation, which ultimately
traces back to the regularization in the ultraviolet,
operating at ultrahigh energies or ultra-small distances
(see, e.g., the discussion in Sec.~V of Ref.~\cite{GhoshKlinkhamer2017}).
Regularization is typically used as
a mere mathematical device, to be removed after the calculation.
But it is also
possible that a particular regularization has a deeper
physical meaning. An example is the lattice
regularization, which could be directly relevant if
spacetime possesses a fundamental discreteness.

Second, the question arises how the anomalous photon term
\eqref{eq:CS-like-term-Abelian} affects
gravity~\cite{Kostelecky2004,Kant2009}. The problem, now, is that
the corresponding term in the energy-momentum tensor is, in general,
neither symmetric nor conserved.
There are solutions of the Einstein equation for special
cases~\cite{Kant2009} but not for the general case.
At this moment, it is not clear how the gravitational theory
can be extended in order to allow for an anomalous
Chern--Simons-like-term in the effective gauge-field action.

\ack
The author thanks M. Schreck for agreeing, on short notice,
to present this talk at the workshop.

\end{document}